\documentstyle[12pt,epsf]{article}
\def \beq{\begin{equation}}
\def \di{\delta_\infty}
\def \eeq{\end{equation}}
\def \mth{M_{\rm th}}
\def \rth{r_{\rm th}}

\begin{document}
\begin{titlepage}
\vspace{-3in}
\rightline{CERN-TH/96-128}
\rightline{DESY-96-089}
\rightline{EFI-96-17}
\rightline{hep-ph/9605373}
\begin{center}

{\large\bf CRITICAL SPACING FOR HEAVY QUARKONIUM DISSOCIATION}\\
\vspace{1.5cm}
{\large Jonathan L. Rosner} \\
\vspace{.5cm}
{\sl CERN, 1211-CH Geneva 23, Switzerland} \\
\vspace{.5cm}
{\sl Deutsches Elektronenen Synchrotron DESY, D-22603 Hamburg, Germany.}
\footnote{Address during initial phase of this work.} \\
\vspace{.5cm}
{\sl Enrico Fermi Institute and Department of Physics}\\
{\sl University of Chicago, Chicago, IL 60637 USA}
\footnote{Permanent address.}\\
\vspace{1.5cm}
\begin{abstract}

When a heavy quark and the corresponding antiquark are separated by more than
$1.4 - 1.5$ fm, it becomes energetically favorable for a light quark-antiquark
pair to be produced, leading to fragmentation into a pair of flavored mesons.
The relation of this critical quark separation to other dimensional constants
of the strong interactions (such as the pion decay constant, the QCD scale, and
the light-quark constituent mass) is discussed. 

\end{abstract} 

\end{center}
\leftline{PACS Codes: 12.39.Pn, 13.25.Gv, 14.40.Gx, 12.38.Gc}
\vfil
\end{titlepage}

The dimensional nature of quantum chromodynamics is manifested in a number of
different ways, all of which are plausibly equivalent to one another.  For
example:  (1) The scale of low-energy pion-pion interactions is set by the
ratios $p_i \cdot p_j/f_\pi^2$, where $p_i$ are pion 4-momenta and $f_\pi =
132$ MeV is the pion decay constant.  (2) A QCD scale $\Lambda_{\rm QCD} =
{\cal O} (200-400)$ MeV whose specific value depends on renormalization
scheme is necessary in order to define the strong-coupling constant $\alpha_s$
at a suitable momentum scale. (3) The masses of $u$ and $d$ quarks as
manifested in hadrons can be regarded as ``constituent-quark'' values $m_q$
of order 300 -- 400 MeV.  The scale of $m_q$, while not as precisely defined
as $f_\pi$ or $\Lambda_{\rm QCD}$, must be related to these two quantities if,
as expected, the limit of zero bare mass for $u$ and $d$ quarks makes sense.

In the present note we add a quantity to this list.  There appears to be a
critical separation between a
heavy quark and a heavy antiquark for which it is energetically favorable to
produce a pair of flavored mesons [as, for example, in $\Upsilon(4S) \to B
\overline{B}$].  This feature was noted in
early studies of the $\Upsilon$ system \cite{EG,QR}. However, the explicit
magnitude of the interquark separation leading to heavy quarkonium dissociation
was not presented.  We find that it is of order $1.4 - 1.5$ fm.  This number is
of current interest for several reasons. 

(1) Lattice gauge theories are approaching a stage where the dimensional
quantities just mentioned (as well as others) can be related to one another.
The study of light-quark production between color centers represented by a
heavy quark and antiquark is becoming feasible \cite{lat} as one learns to cope
with the ``unquenched'' version of QCD in which light quark-antiquark pairs are
properly treated. 

(2) The quarkonium systems to which the critical dissociation distance applies
include $c \bar c$ and $b \bar b$ systems, for which there remain prospects for
discovering a few more P-wave and D-wave states, and $b \bar c$ states, for
which there are extensive theoretical studies of the spectroscopy \cite{EQ} and
a hint of the ground state \cite{bchint}.  The top quark is too heavy to have a
quarkonium spectroscopy (since its decay width will be greater than the
expected level spacing), but the critical-separation parameter should still
apply to $t \bar t$ pairs or their products $t \bar b$, $\bar t b$, or $b \bar
b$ \cite{OR}. 

(3) One might expect the critical separation of a pair of static color centers
for dissociation into a pair of flavored mesons to be a fundamental parameter
in theories \cite{HQCPT} linking heavy-quark physics and chiral perturbation
theory.

We assume that a flavor-independent potential $V(r)$ describes all bound states
$Q \bar Q$ of a heavy quark $Q$ and its antiquark as well as flavored mesons $Q
\bar q$ and $\bar Q q$ involving the heavy quark $Q$ and a light quark $q =
u,~d$.  In the notation of Ref.~\cite{QR}, we define $\delta(m_Q) \equiv
2m({\rm lowest}~Q \bar q) - 2 m_Q$.  This quantity should tend to a finite
limit $\di$ as $m_Q \to \infty$. The quantity $\di/2$ is the same as the
parameter $\bar \Lambda$ of heavy quark effective theory \cite{HQET}. We seek
the value of the critical
threshold $Q \bar Q$ separation $\rth$ for which $\di =
V(\rth)$. A trivial modification permits the description of bound states such
as $b \bar c$ involving unlike-mass quarks. 

The value of $\rth \equiv V^{-1}(\di)$ will be universal to the extent that (a)
$\di$ is well-defined (as expected in heavy quark effective theory), and (b) a
flavor-independent potential $V(r)$ actually provides a good description of the
$Q \bar Q$ interaction near flavor threshold.  The parameters $\di/2$ and the
light-quark constituent mass $m_q$ are closely related; neither parameter can
be chosen arbitrarily in a theory with given $f_\pi$ or $\Lambda_{\rm QCD}$.
The assumption of a flavor-independent potential has been reasonably well borne
out \cite{QRU,BT} by comparison of $c \bar c$ and $b \bar b$ systems, and will
be tested further in studies of $b \bar c$ states \cite{EQ}. 
 
We calculate the critical separation using a simple phenomenological potential
\cite{logrefs}.  We then compare it with values obtained in a more
model-independent manner, and discuss its relation to other dimensional
constants in QCD. 

A satisfactory interpolation between charmonium ($c \bar c$) and upsilon ($b
\bar b$) states is provided by a potential of the form $V(r) = C \ln (r/r_0)$,
with $C \simeq 0.72$ GeV and $r_0$ depending upon the specific choice of $c$
and $b$ quark masses \cite{logrefs}.  The Schr\"odinger equation for the
reduced radial wave function $u(r)$ of S-wave bound states is
\beq \label{eqn:schr}
- \frac{1}{2 \mu} \frac{d^2u}{dr^2} + C \ln \left( \frac{r}{r_0} \right) u
= E u~~~,
\eeq
where $\mu$ is the reduced mass: $2 \mu = m_Q$ for a $Q \bar Q$ bound state,
and $E = M(Q \bar Q) - 2 m_Q$. In terms of dimensionless variables $\epsilon
\equiv E/C$ and $\rho \equiv r \sqrt{2 \mu C}$, (\ref{eqn:schr}) becomes
\beq
- \frac{d^2u}{d\rho^2} + \ln \rho u = [\epsilon + \ln(r_0 \sqrt{2 \mu C})]u~~~.
\eeq
The lowest eigenvalue of this equation is $\epsilon + \ln(r_0 \sqrt{2 \mu C}) =
1.0443$, based on a numerical solution \cite{QMQ}.  Consequently, one may
eliminate the parameter $r_0$ in favor of the ground state mass, the
parameter $C$, and the reduced mass.  Recalling the definition of the
radius $\rth$ and setting $\mth - 2m_Q = V(\rth)$, we finally have
\beq
[\mth - M(1S)]/C + 1.0443 = \ln(\rth\sqrt{2 \mu C})~~~.
\eeq

We begin by neglecting spin-dependent effects in both $Q \bar Q$ and $Q \bar q$
systems, and reduced-mass effects in the $Q \bar q$ system.  This approximation
is most reliable for the $\Upsilon$ levels.  Consequently, taking $\mth = 2
M(B) = 10.558$ GeV, $M(1S) = M[\Upsilon(1S)] = 9.460$ GeV, $C = 0.72$ GeV, and
a range of $2 \mu = m_b$ between 4.5 and 5 GeV, we find the values of $\rth$
shown as the dashed line in Fig.~1. 

To account for the hyperfine splittings in the $Q \bar Q$ and $Q \bar q$
systems and the reduced-mass effect in the $Q \bar q$ system, we assume that
the $\Upsilon(1^3S_1)$ level is 10 MeV above the spin-averaged $1S$ mass
\cite{NPT}.  Furthermore, we apply the corrections \cite{QR} 
\beq
\delta(m_Q)_{\rm hfs} = \frac{m_b}{m_Q} \cdot \frac{3}{2} (m_{B^*} - m_B)
= 69~{\rm MeV} \left( \frac{m_b}{m_Q} \right)
\eeq
and
\beq
\delta(m_Q)_{\rm red.~mass} = - C \ln[\mu(m_Q)/\mu(m_b)] \simeq
C m_q \left( \frac{1}{m_Q} - \frac{1}{m_b} \right)
\eeq
to estimate
\beq
\di - \delta(m_b) \simeq 10~{\rm MeV} + 69~{\rm MeV} - 46~{\rm MeV}
= 33~{\rm MeV}~~~,
\eeq
where the first term is our estimate of the $\Upsilon (1S)$ hyperfine term, the
second is the contribution of the hyperfine term in $2 M(B)$, and the third is
the reduced-mass effect, estimated for the logarithmic potential.  One is thus
assuming such a potential to hold not only for $Q \bar Q$ but also for $Q \bar
q$ states.  We have used a light-quark constituent mass \cite{QM} $m_q = 310$
MeV.  With the above corrections, we now estimate the range of $\rth$ shown as
the solid line in Fig.~1. The corrected values of $\rth$ range between about
7.6 and 7.2 GeV$^{-1}$ (1.5 -- 1.42 fm) for $4.5 \le m_b \le 5$ GeV. 

For comparison, the dot-dashed line in Fig.~1 depicts the values of $\rth$
obtained from an inverse-scattering construction of the interquark potential
\cite{QRU} using the $\Upsilon$ levels. These values have not been corrected
for hyperfine or reduced-mass effects, so they should be compared with the
uncorrected values obtained above.  A power-law potential \cite{GRR} fitting
charmonium and $\Upsilon$ spectra with $(m_c,m_b) = (1.56,4.96)$ GeV gives
$\rth = 7.2$ GeV$^{-1}$ (uncorrected) and 7.6 GeV$^{-1}$ (corrected). The
agreement of the various estimates is fairly good, indicating that
model-dependent effects are unlikely to affect the estimate significantly. As
shown in inverse-scattering \cite{QRU} and explicit potential \cite{BT}
calculations, the shape of any smooth potential which reproduces a given set of
energy levels is fairly well specified for the range of energies corresponding
to the known levels.  Once a potential is required to reproduce the
$\Upsilon(1S-4S)$ levels and their leptonic widths, that potential's shape is
specified between roughly 0.1 fm = 0.5 GeV$^{-1}$ and the interquark separation
corresponding to $V(r) \approx M[\Upsilon(4S)] - 2 m_b$, which is just above
flavor threshold.

\begin{figure}
\centerline{\epsfysize=4in \epsffile{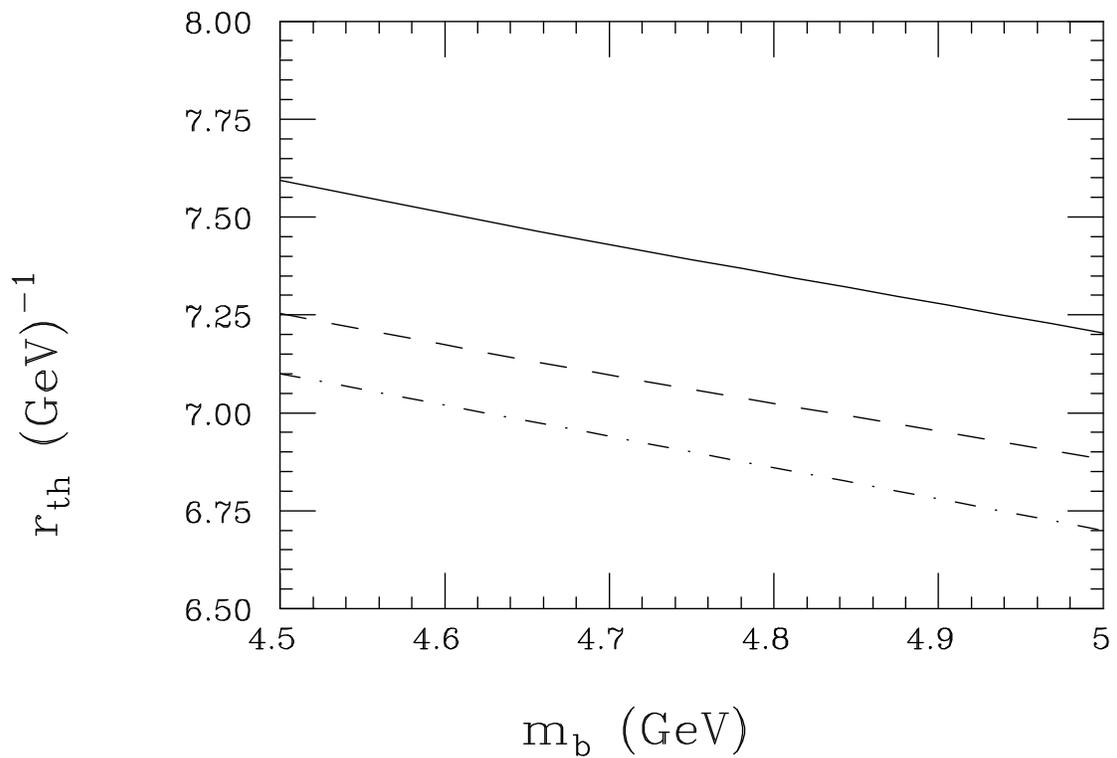}}
\caption{Dependence of threshold parameter $\rth$, as estimated from
$\Upsilon$ levels, on assumed value $m_b$ of bottom quark mass.  Dashed
line:  uncorrected values assuming logarithmic potential.  Solid line:
values assuming logarithmic potential corrected for hyperfine and
reduced-mass effects.  Dash-dotted line:  uncorrected values obtained
from an inverse-scattering approach.}
\end{figure}

Lattice gauge theories are able to estimate the distance at which a string
of chromoelectric flux breaks by the deviation from a linear behavior of
the field energy as a function of interquark separation.  It is not clear that
such a distance corresponds to flavor threshold, since discrete $b \bar b$
resonances (and indeed, series of light-quark resonances) persist well above
flavor threshold.  Nonetheless, our estimate of $\rth$ is in accord with an
upper bound of $1.9 \pm 0.2 \pm 0.2$ fm \cite{latq} (quoted in Ref.~\cite{lat}
as 1.7 fm on the basis of a calculation in Ref.~\cite{Sommer}) for the breaking
of a QCD string obtained using a quenched approximation in lattice gauge
theory.  In a quenched lattice of size (1.5 fm)$^3$, no breaking of a QCD
string has been observed; the linear behavior persists out to the maximum
accessible interquark separation.  It would be interesting to see whether at
slightly larger distances a linear behavior could actually {\it coexist} with
production of light quark-antiquark pairs. 

How might the critical spacing parameter $\rth$ be related to other
dimensionful quantities in QCD?  It is clearly related to the constituent-quark
mass since a certain amount of chromoelectric energy is required to create the
$q \bar q$ pair.  In turn, the constituent-quark mass scale is set
\cite{QM,DGG} by the need to agree with such quantities as $m_\rho$ and $m_p$,
whose values are related to the QCD scale $\Lambda_{\rm QCD}$. It has been
argued \cite{BG} (cf., however, Refs.~\cite{VV}) that the mass scale of
resonances like the $\rho$ meson can be regarded as a number of order $2 \pi
f_\pi$ in a chirally symmetric theory involving massless pions. In any event a
direct relation between $f_\pi$ and $\Lambda_{QCD}$ in a theory of massless
pions is highly likely. 

Further dimensionful quantities in the strong interactions which might bear a
relation to those mentioned include the universal string tension describing the
long-distance interquark interaction $V(r) \simeq k r$, $k \simeq 0.18$
GeV$^2$, and the universal slope $\alpha'$ of Regge trajectories for
light-quark systems, $\alpha' = 1/(2 \pi k) \simeq 0.88$ GeV$^{-2}$ \cite{YN}. 
The production of light quark-antiquark pairs in a linear potential has been
considered some time ago \cite{CN}. 

To conclude, we have argued that once a heavy color triplet and antitriplet 
become separated by more than 1.4 -- 1.5 fm, the chromoelectric flux lines
joining them contain sufficient energy to produce a light quark-antiquark
pair, leading to the decay of the heavy quarkonium system into a pair of
flavored mesons.  It would be interesting to see whether current lattice
gauge and chiral theories of non-perturbative QCD could relate this quantity
to others which set the QCD scale.

\section*{Acknowledgments}

I thank the CERN and DESY Theory Groups for their hospitality during this work,
and W. Buchm\"uller and M. L\"uscher for fruitful discussions. This work was
supported in part by the United States Department of Energy under Contract
No.~DE FG02 90ER40560. 
 
\def \ajp#1#2#3{Am.~J.~Phys.~{\bf#1} (#3) #2}
\def \apas#1#2#3{Acta Phys.~Austriaca Suppl.~{\bf#1} (#3) #2}
\def \apny#1#2#3{Ann.~Phys.~(N.Y.) {\bf#1} (#3) #2}
\def \app#1#2#3{Acta Phys.~Polonica {\bf#1} (#3) #2}
\def \arnps#1#2#3{Ann.~Rev.~Nucl.~Part.~Sci.~{\bf#1} (#3) #2}
\def \cmp#1#2#3{Commun.~Math.~Phys.~{\bf#1} (#3) #2}
\def \cmts#1#2#3{Comments on Nucl.~Part.~Phys.~{\bf#1} (#3) #2}
\def \cn{Collaboration}
\def \corn93{{\it Lepton and Photon Interactions:  XVI International Symposium,
Ithaca, NY August 1993}, AIP Conference Proceedings No.~302, ed.~by P. Drell
and D. Rubin (AIP, New York, 1994)}
\def \cp89{{\it CP Violation,} edited by C. Jarlskog (World Scientific,
Singapore, 1989)}
\def \dpff{{\it The Fermilab Meeting -- DPF 92} (7th Meeting of the American
Physical Society Division of Particles and Fields), 10--14 November 1992,
ed. by C. H. Albright \ite~(World Scientific, Singapore, 1993)}
\def \dpf94{DPF 94 Meeting, Albuquerque, NM, Aug.~2--6, 1994}
\def \efi{Enrico Fermi Institute Report No. EFI}
\def \el#1#2#3{Europhys.~Lett.~{\bf#1} (#3) #2}
\def \f79{{\it Proceedings of the 1979 International Symposium on Lepton and
Photon Interactions at High Energies,} Fermilab, August 23-29, 1979, ed.~by
T. B. W. Kirk and H. D. I. Abarbanel (Fermi National Accelerator Laboratory,
Batavia, IL, 1979}
\def \hb87{{\it Proceeding of the 1987 International Symposium on Lepton and
Photon Interactions at High Energies,} Hamburg, 1987, ed.~by W. Bartel
and R. R\"uckl (Nucl. Phys. B, Proc. Suppl., vol. 3) (North-Holland,
Amsterdam, 1988)}
\def \ib{{\it ibid.}~}
\def \ibj#1#2#3{~{\bf#1} (#3) #2}
\def \ichep72{{\it Proceedings of the XVI International Conference on High
Energy Physics}, Chicago and Batavia, Illinois, Sept. 6--13, 1972,
edited by J. D. Jackson, A. Roberts, and R. Donaldson (Fermilab, Batavia,
IL, 1972)}
\def \ijmpa#1#2#3{Int.~J.~Mod.~Phys.~A {\bf#1} (#3) #2}
\def \ite{{\it et al.}}
\def \jmp#1#2#3{J.~Math.~Phys.~{\bf#1} (#3) #2}
\def \jpg#1#2#3{J.~Phys.~G {\bf#1} (#3) #2}
\def \lkl87{{\it Selected Topics in Electroweak Interactions} (Proceedings of
the Second Lake Louise Institute on New Frontiers in Particle Physics, 15--21
February, 1987), edited by J. M. Cameron \ite~(World Scientific, Singapore,
1987)}
\def \ky85{{\it Proceedings of the International Symposium on Lepton and
Photon Interactions at High Energy,} Kyoto, Aug.~19-24, 1985, edited by M.
Konuma and K. Takahashi (Kyoto Univ., Kyoto, 1985)}
\def \mpla#1#2#3{Mod.~Phys.~Lett.~A {\bf#1} (#3) #2}
\def \nc#1#2#3{Nuovo Cim.~{\bf#1} (#3) #2}
\def \np#1#2#3{Nucl.~Phys.~{\bf#1} (#3) #2}
\def \pisma#1#2#3#4{Pis'ma Zh.~Eksp.~Teor.~Fiz.~{\bf#1} (#3) #2 [JETP Lett.
{\bf#1} (#3) #4]}
\def \pl#1#2#3{Phys.~Lett.~{\bf#1} (#3) #2}
\def \plb#1#2#3{Phys.~Lett.~B {\bf#1} (#3) #2}
\def \pr#1#2#3{Phys.~Rev.~{\bf#1} (#3) #2}
\def \pra#1#2#3{Phys.~Rev.~A {\bf#1} (#3) #2}
\def \prd#1#2#3{Phys.~Rev.~D {\bf#1} (#3) #2}
\def \prl#1#2#3{Phys.~Rev.~Lett.~{\bf#1} (#3) #2}
\def \prp#1#2#3{Phys.~Rep.~{\bf#1} (#3) #2}
\def \ptp#1#2#3{Prog.~Theor.~Phys.~{\bf#1} (#3) #2}
\def \rmp#1#2#3{Rev.~Mod.~Phys.~{\bf#1} (#3) #2}
\def \rp#1{~~~~~\ldots\ldots{\rm rp~}{#1}~~~~~}
\def \si90{25th International Conference on High Energy Physics, Singapore,
Aug. 2-8, 1990}
\def \slc87{{\it Proceedings of the Salt Lake City Meeting} (Division of
Particles and Fields, American Physical Society, Salt Lake City, Utah, 1987),
ed.~by C. DeTar and J. S. Ball (World Scientific, Singapore, 1987)}
\def \slac89{{\it Proceedings of the XIVth International Symposium on
Lepton and Photon Interactions,} Stanford, California, 1989, edited by M.
Riordan (World Scientific, Singapore, 1990)}
\def \smass82{{\it Proceedings of the 1982 DPF Summer Study on Elementary
Particle Physics and Future Facilities}, Snowmass, Colorado, edited by R.
Donaldson, R. Gustafson, and F. Paige (World Scientific, Singapore, 1982)}
\def \smass90{{\it Research Directions for the Decade} (Proceedings of the
1990 Summer Study on High Energy Physics, June 25 -- July 13, Snowmass,
Colorado), edited by E. L. Berger (World Scientific, Singapore, 1992)}
\def \smassb{{\it Proceedings of the Workshop in $B$ Physics at Hadron
Colliders} (Snowmass, CO, June 21 -- July 2, 1993), edited by P. McBride
and C. S. Mishra, Fermilab report Fermilab-CONF-93/267 (1993)}
\def \stone{{\it B Decays}, edited by S. Stone (World Scientific, Singapore,
1994)}
\def \tasi90{{\it Testing the Standard Model} (Proceedings of the 1990
Theoretical Advanced Study Institute in Elementary Particle Physics, Boulder,
Colorado, 3--27 June, 1990), edited by M. Cveti\v{c} and P. Langacker
(World Scientific, Singapore, 1991)}
\def \yaf#1#2#3#4{Yad.~Fiz.~{\bf#1} (#3) #2 [Sov.~J.~Nucl.~Phys.~{\bf #1} (#3)
#4]}
\def \zhetf#1#2#3#4#5#6{Zh.~Eksp.~Teor.~Fiz.~{\bf #1} (#3) #2 [Sov.~Phys. -
JETP {\bf #4} (#6) #5]}
\def \zpc#1#2#3{Zeit.~Phys.~C {\bf#1} (#3) #2}

\end{document}